\documentclass[%
 reprint,
showpacs,
amsmath,amssymb,
aps,
]{revtex4-1}

\usepackage[breaklinks=true,colorlinks=true,linkcolor=blue,urlcolor=blue,citecolor=blue]{hyperref}
\usepackage{graphicx}
\usepackage{dcolumn}
\usepackage{bm}

\begin{document}

\title{Noise driven current reversal and stabilisation in the tilted ratchet potential subject to tempered stable L{\'e}vy noise }

\author{Mathew L. Zuparic}
\affiliation{Defence Science and Technology Group, Canberra, 2600, Australia}

\author{Alexander C. Kalloniatis}
\affiliation{Defence Science and Technology Group, Canberra, 2600, Australia}

\author{Dale O. Roberts}
\affiliation{Australian National University, ACT 2601, Australia}

\date{\today}


\begin{abstract}
We consider motion of a particle in a one-dimensional tilted ratchet potential subject to
two-sided tempered stable L{\'e}vy noise characterised by strength $\Omega$, fractional index $\alpha$,
skew $\theta$ and tempering $\lambda$. We derive analytic solutions to the corresponding Fokker-Planck L\'{e}vy equations
for the probability density. Due to the periodicity of the potential, we carry out reduction to a compact domain
and solve for the analogue there of steady-state solutions which we represent as wrapped probability density functions.
By solving for the expected value of the current associated with the particle motion, we are able to determine threshold
for metastability of the system, namely when the particle stabilises in a well of the potential and when the particle is in motion,
for example as a consequence of the tilt of the potential. Because the noise may be asymmetric, we examine the relationship between
skew of the noise and the tilt of the potential. With tempering, we find two remarkable regimes where the current may be reversed in a direction opposite to the tilt or where the particle may be stabilised in a well in circumstances where deterministically it should flow with the tilt.
\end{abstract}

\maketitle

\vspace{2pc}

\section{Introduction}
\label{intro}

The impact of stochastic driving in non-linear systems \cite{Reimann02} is significant in diverse physical, chemical, and biological  systems; we would
also include here social systems for reasons to be given soon. For all the sophistication of many models here, some features are common: a potential
energy landscape with a tilt -
or bias - superimposed with some form
of periodicity in the distributions of local minima. These are known as tilted ratchet potentials because of the behaviour that, under metastable circumstances, a particle may 
drift down the tilt but temporarily `catch' in the periodically occuring wells \cite{Linder01}.
Many works have addressed Gaussian or Brownian noise in such potentials, or analogous forms such as `washboard' potentials
\cite{Henn2009,MulHenn2011}, and multi-dimensional forms of these \cite{ChallJack2013}. However, more general forms of noise particularly with jumps drawn from heavy-tails \cite{BorovkovBorovkov08} are now topical in the literature, for example spatially
stable L{\'e}vy noise \cite{Chambers76,Twee1984} and tempered stable  L{\'e}vy noise \cite{MantStan1994,Koponen1995,BaeMeer2010,GajMag2010,MeerSab2016} -  leading to the study of so-called fractional (or super) diffusion
\cite{MetzKlaft2000,Jumarie2004,MetzKlaft2004}. There is a long history of applications to more standard one-dimensional systems
\cite{JMF1999,MetzKlaft2000b,CGKMT2002,MagWer2007,CGGKS2008,Gorska2011}.
Such non-Gaussian noise models have been applied to ratchet potentials in the absence of tilt \cite{KullCast2012}. In this paper we extend these analyses to the case of non-zero tilt.

The signature behaviours of interest in stochastic tilted ratchets \cite{RPHR2002,Henn2009,MulHenn2011} are those of current stabilisation - where the particle may be localised in a well under
the influence of the noise when deterministically it would seek to roll down the tilt - and current {\it reversal} - where the particle
is driven against the tilt by the noise. This phenomena form a subset of non-linear collective behaviours alongside
stochastically driven resonance \cite{GHJM1998,GHJM2009} and synchronisation \cite{Kawa2014}. 
Given the importance of noise characterised by jumps and heavy tails in diverse contexts \cite{Kleinert2009} such as the fluid properties of
plasmas \cite{delCastCarrLyn2005,delCastGonCheck2008}, finance \cite{EllvdHoek2003,CarteaDelCast2007} and brain waves
\cite{RobBoonBreak2015}, the application of this to the collective phenomena of tilted ratchets is evident.
Already in the presence of periodic potentials under tempered stable noise, current reversal has been observed \cite{KullCast2012}. 
In this paper we explore the role in this phenomenon of the tilt - which should naturally inhibit current reversal - in the vicinity
of the deterministic threshold for current flow.

Our interest in this problem arises from a quite different problem, that of synchronisation on networks as 
exemplified in the stylised Kuramoto model \cite{Kur84}. Firstly, the property of metastability for the Kuramoto model on
ring graphs subject to weak Gaussian noise has been identified \cite{DeVille12}. This uses the approach
to stochastic metastability of Freidlin-Wentzell (FW) theory \cite{Freidlin98}. Secondly, the Kuramoto model close to the synchronisation threshold
maps to a tilted ratchet potential; some of us have explored this property for a generalisation of the Kuramoto model for two populations
on separate networks \cite{ZupKall2013} and subject to Gaussian noise \cite{Holder2017}. Finally, some of us have begun explorations of the ordinary Kuramoto model
subject to stable \cite{KallRob2017} and tempered stable noise \cite{KallRob2017b} - with hints at stochastic synchronisation there as well.

The paper is structured as follows. In the next section we set-up the tempered stable stochastic system, first using the fractional Langevin 
formalism and then the Fokker-Planck formalism. We then discuss the reduction procedure in light of the periodic structure of the
potential, using the Gaussian case as an example. We then present the main results showing the solution to the
reduced Fokker-Planck probability density and associated expected value of the current associated with the particle; in this
we use a little-known representation of the solutions as `wrapped probability densities'. We examine
these under variations of parameters such as the fractional $\alpha$, the tilt and the tempering of the noise. Here we identify
the regimes of current reversal and stabilisation.
The paper
concludes with prospects for future work.

\section{Tempered-fractional-Fokker-Planck equations}

\subsection{Fractional Langevin equations}
We begin with a generalisation of the Langevin equation capturing the combination of deterministic and stochastic dynamics written in
stochastic differential notation:
\begin{equation}
dx(t)=-V'(x(t)) dt + dL^{\alpha,\theta,\lambda}(t)
\label{defining-Lang}
\end{equation}
where $\alpha \in (0,1) \cup (1,2]$ is the fractional index describing the
power law behaviour of the L{\'e}vy process, $\theta \in [-1,1]$ is the asymmetry or skew of the noise distribution, and $\lambda \in (0,\infty)$ is the tempering parameter
for the heavy tails \cite{KawMas2011}. For $\alpha = 2$, the L\'{e}vy noise term in Eq.(\ref{defining-Lang}) becomes Gaussian.

For this work the potential $V(x)$ is given by
\begin{equation}
V(x) = -\mu x - \gamma \cos(x-\rho), \;\; \rho \in (-\pi,\pi]
\label{ratchetpot}
\end{equation}
where $\mu \in \mathbb{R}$ and $\gamma \in \mathbb{R}$ are referred to as the \textit{tilt} and \textit{amplitude} respectively. Eq.(\ref{ratchetpot}) is commonly referred to as a \textit{tilted periodic ratchet} \cite{Linder01,Reimann02}. The potential has two important features. The first is the periodicity $V'(x)=V'(x+2\pi)$. The second is given by the sign of the quantity
\begin{equation}
{\cal K} = \gamma^2 - \mu^2
\label{specK}
\end{equation}
which encapsulates the interplay between the tilt and the amplitude. Specifically, if ${\cal K}>0$ then $V(x)$ contains a series of local minima (\textit{stable fixed points}) at $x=\rho + \sin^{-1}(\mu/\gamma)$ mod $2\pi$, and a series of local maxima (\textit{unstable fixed points}) at $x=\pi + \rho - \sin^{-1}(\mu/\gamma)$ mod $2\pi$. If ${\cal K}=0$ then the maxima and minima collapse into each other and form a series of inflection points (\textit{unstable fixed points}). Lastly, if ${\cal K} <0$ then $V(x)$ becomes an entirely monotonic function with no inflection points, namely no stationary points.

\subsection{Tempered-fractional diffusion}
The probability density ${\cal P}(x,t)$ associated with the tempered-stable L\'{e}vy process in Eq.(\ref{defining-Lang}) obeys the following 
so-called Tempered Fractional Fokker-Planck equation (TFFP) \cite{Cartea07}
\begin{eqnarray}
\begin{split}
\frac{\partial}{\partial t} {\cal P}(x,t) = \left\{ \Omega \partial^{\alpha,\theta,\lambda}_x + \frac{\partial}{\partial x} V'(x)  \right\} {\cal P}(x,t)\\
{\cal P} (x,0) = \delta(x-y),
\end{split}
\label{defining-FP}
\end{eqnarray}
where $\Omega \in \mathbb{R}_+$ is the diffusivity and the operator $\partial^{\alpha,\theta,\lambda}_x$ is the tempered-fractional-diffusion operator, given explicitly by \cite{KullCast2012}
\begin{equation}
\partial^{\alpha,\theta,\lambda}_x = {\cal D}^{\alpha,\theta,\lambda}_x + v^{\alpha,\theta,\lambda} \frac{\partial}{\partial x} + \nu^{\alpha,\lambda}.
\end{equation}
Here $v^{\alpha,\theta,\lambda}$ and $\nu^{\alpha,\lambda}$ are so-called `induced' drift and source/sink terms given by,
\begin{equation}
v^{\alpha,\theta,\lambda}= \left\{  \begin{array}{cl}
0, & \alpha \in (0,1)\\
\frac{\alpha \theta \lambda^{\alpha-1}}{|\cos(\pi \alpha/2)|} & \alpha \in (1,2)
\end{array}
\right. , \;\; \nu^{\alpha,\lambda} =  \frac{\lambda^{\alpha}}{\cos(\pi \alpha /2)}.
\end{equation}
The operator ${\cal D}^{\alpha,\theta,\lambda}_x$ is the $\lambda$-truncated fractional derivative of order $\alpha$, given by,
\begin{equation}
{\cal D}^{\alpha,\theta,\lambda}_x = l(\theta) e^{-\lambda x} \,_{-\infty} D^{\alpha}_x e^{\lambda x} - r(\theta) e^{\lambda x} \,_x D^{\alpha}_{\infty} e^{-\lambda x}
\end{equation}
where the operators $ \,_{-\infty} D^{\alpha}_x$ and $\,_x D^{\alpha}_{\infty}$ are the Riemann-Liouville derivatives \cite{del-Castillo-Negrete12}. Both operators have the following form in  Fourier space \cite{Podlubny99,Samko93}
\begin{eqnarray}
\begin{split}
{\cal F} \left[e^{-\lambda x} \,_{-\infty} D^{\alpha}_x e^{\lambda x} f(x) \right] = (\lambda-i k)^{\alpha} \widehat{f}(k)\\
{\cal F} \left[e^{\lambda x} \,_x D^{\alpha}_{\infty} e^{-\lambda x} f(x) \right] = (\lambda+i k)^{\alpha} \widehat{f}(k)
\end{split}
\end{eqnarray}
where our convention for the Fourier transform is:
\begin{eqnarray}
\begin{split}
{\cal F} \left[ f(x) \right] = \int^{\infty}_{-\infty}dx e^{i k x}f(x) =  \widehat{f}(k)\\
{\cal F}^{-1} \left[ \widehat{f}(k) \right] = \int^{\infty}_{-\infty} \frac{dk}{2\pi} e^{-i k x}\widehat{f}(k) =  f(x).
\end{split}
\end{eqnarray}
Finally, for definitional purposes, the weighting factors
\begin{equation}
l(\theta) = \frac{\theta-1}{2 \cos(\pi \alpha/2)}, \;\; r(\theta) = \frac{\theta+1}{2 \cos(\pi \alpha/2)}
\end{equation}
give the asymmetry imposed on each of the Riemann-Liouville derivatives. Taking the Fourier transform of the fractional derivative $ \partial^{\alpha,\theta,\lambda}_x$ we obtain
\begin{eqnarray}
\begin{split}
{\cal F} \left[  \partial^{\alpha,\theta,\lambda}_x  f(x) \right] = \left\{l(\theta) (\lambda-i k)^{\alpha}  \right.\\
\left.-  r(\theta)(\lambda+i k)^{\alpha} - i k v^{\alpha,\theta,\lambda} +  \nu^{\alpha,\lambda} \right\} \widehat{f}(k)\\
= \Lambda(k) \widehat{f}(k)
\end{split}
\label{lambda}
\end{eqnarray}
where $\Lambda(k)$ is the logarithm of the characteristic function of the tempered stable L\'{e}vy process.
For example, the limit $\lambda\rightarrow 0$ reproduces the L{\'e}vy-Khinchine law for the stable process \cite{Sato99}.

\subsection{Illustrative example - Gaussian limit}
In order to proceed with the general TFFP equation with the tilted ratchet potential, we first briefly detail the Gaussian limit case, namely
\begin{equation}
\frac{\partial}{\partial t} {\cal P}(x,t) = \left\{ \Omega \frac{\partial^2}{\partial x^2} + \frac{\partial}{\partial x} V'(x)  \right\} {\cal P}(x,t)
\label{Gauss}
\end{equation}
as its relatively simple explanation of solution allows for greater intuition when we consider the more complicated fractional case. As well as the defining equation, Eq. (\ref{Gauss}), we are also concerned with the \textit{probability current} ${\cal J}(x,t)$, defined by the following probability conservation expression
\begin{eqnarray}
\begin{split}
&&\frac{\partial}{\partial t} {\cal P}(x,t) +\frac{\partial}{\partial x} {\cal J}(x,t) = 0\\
\Rightarrow && {\cal J}(x,t) = - \left\{    \Omega \frac{\partial}{\partial x} +  V'(x)  \right\} {\cal P}(x,t).
\end{split}
\label{currentGauss}
\end{eqnarray}
As explained in \cite{Reimann02} and references therein, one finds a non-normalisable density when following the usual procedure for a
steady-state solution:
solving the Pearson equation for the steady state density ${\cal P}_{st}(x)$, namely setting $\frac{\partial}{\partial t} {\cal P}(x,t) = 0$ in Eq.(\ref{Gauss}), and applying the vanishing boundary condition ${\cal P}_{st}(x) \rightarrow 0$ at the natural boundaries $x \rightarrow \pm \infty$. This is due to the
periodicity of the ratchet potential and its \textit{metastability} in the presence of noise \cite{Freidlin98,Berglund07,DeVille12}. In order to circumvent this phenomenon, we restrict the support of $x$ to $\mathbb{S}^1$ by constructing the so-called \textit{reduced density} ${\cal P}^{(r)}(x,t)$ and \textit{reduced probability current} ${\cal J}^{(r)}(x,t)$ through
\begin{eqnarray}
\begin{split}
{\cal P}^{(r)}(x,t) \equiv \sum^{\infty}_{n=-\infty} {\cal P}(x+2 \pi n,t), \\
 {\cal J}^{(r)}(x,t) \equiv \sum^{\infty}_{n=-\infty} {\cal J}(x+2 \pi n,t).
\end{split}
\label{reduced}
\end{eqnarray}
Due to the linearity of the Fokker-Planck equation, the reduced density and reduced probability current also obey Eqs.(\ref{Gauss}) and (\ref{currentGauss}) respectively, but with the new boundary and normalisation conditions
\begin{equation}
{\cal P}^{(r)}(-\pi,t) = {\cal P}^{(r)}(\pi,t), \;\; \int^{\pi}_{-\pi} dx{\cal P}^{(r)}(x,t) = 1.
\label{boundandnorm}
\end{equation} 
The reduced steady state density is then given by \cite{Stratonovich67}
\begin{equation}
{\cal P}_{st}^{(r)}(x) = \frac{\sinh \left( \frac{\pi \mu}{\Omega} \right) e^{-\frac{V(x)}{\Omega}}\int^{x+2\pi}_{x}d\varphi e^{\frac{V(\varphi)}{\Omega}}}{2 \pi^2 \left| I_{i \mu} \left( \frac{\gamma}{\Omega} \right) \right|^2 \left( 1-e^{-\frac{2 \pi \mu}{\Omega}} \right)} 
\label{PDFGauss}
\end{equation}
where $I_{i \mu}$ is the modified Bessel function of imaginary order.
\begin{figure*}
\includegraphics[height=7cm]{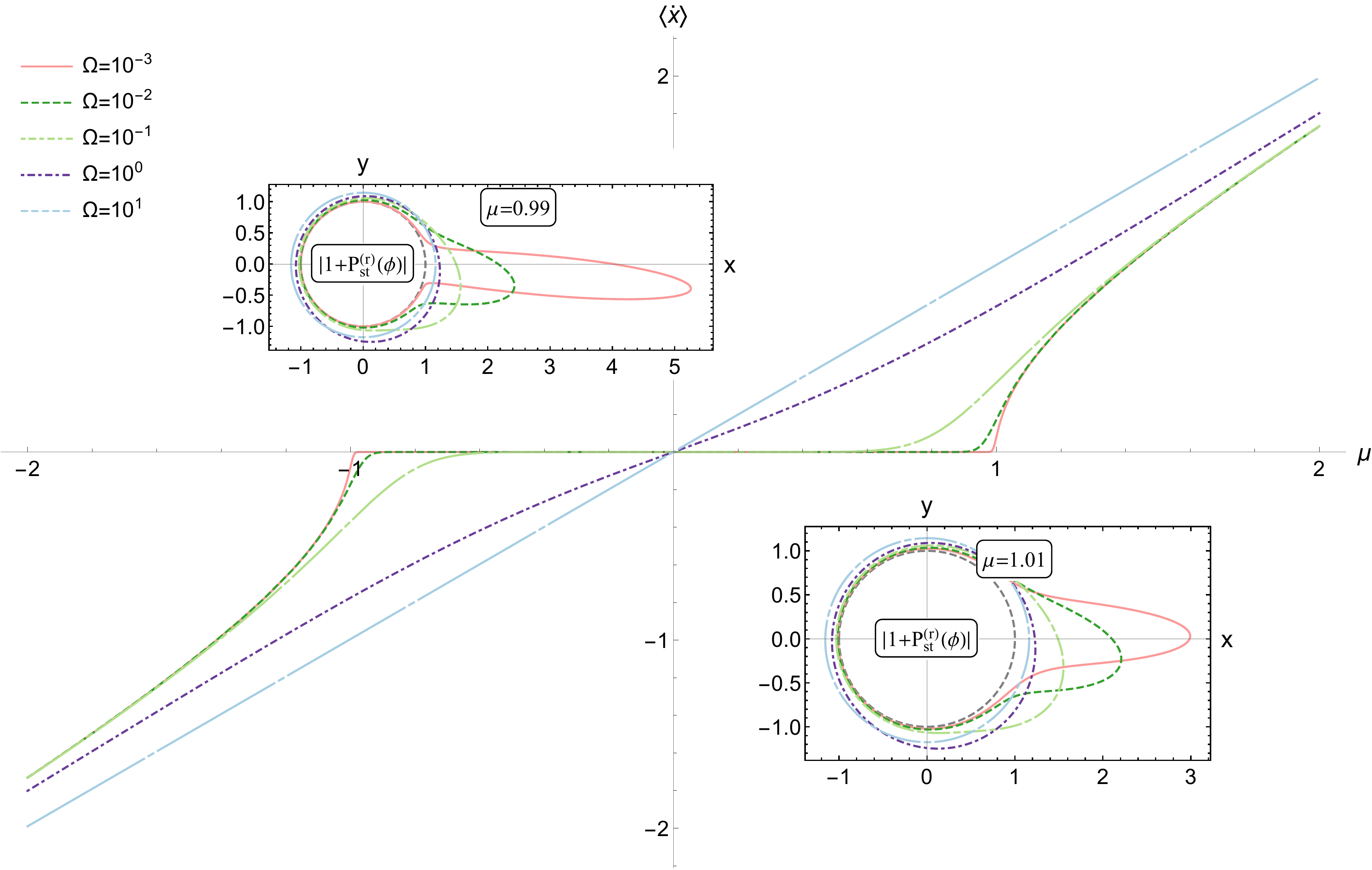}
\caption{Examples of Eq.(\ref{PDFGauss}) (insets) and Eq.(\ref{aveGauss}) (main) for $\rho=-1.5$ and $\gamma=1$ and various values of the diffusivity $\Omega$ which governs noise strength. The leftmost inset details a case with stable fixed points in the potential ($\mu=0.99$), the rightmost inset details a case with no fixed points in the potential ($\mu=1.01$). Both insets are given as a parametric plot governed by Eq.(\ref{parametric}). }
\label{fig:gaussPDF}
\end{figure*}

Additionally, the corresponding average velocity $\langle \dot{x} \rangle$, which may assume non-zero values
as a consequence of the metastability and the tilt of the potential $\mu$, is given by \cite{Stratonovich67}
\begin{equation}
\langle \dot{x} \rangle = \frac{d}{dt}\langle x \rangle = \int^{\pi}_{-\pi} dx{\cal J}^{(r)}_{st}(x) = \frac{\Omega \sinh \left( \frac{\pi \mu}{\Omega} \right)}{ \pi \left| I_{i \mu} \left( \frac{\gamma}{\Omega} \right) \right|^2}.
\label{aveGauss}
\end{equation}
Note that, trivially, for Gaussian noise there is zero average velocity in the absence of tilt: $\langle \dot{x} \rangle =0$ if $\mu=0$.

Fig.\ref{fig:gaussPDF} gives examples of Eq.(\ref{PDFGauss}) (insets) and Eq.(\ref{aveGauss}) (main figure)
where we have fixed the amplitude $\gamma=1$; thus deterministically, the point of instability ${\cal K}=0$ occurs when $\mu=\pm1$
from Eq.(\ref{specK}) so for $\mu=(-1,1)$ $\dot{x}(t\rightarrow\infty) =0$. We also choose the offset $\rho=-1.5$.
The main part of Fig. \ref{fig:gaussPDF} shows the average velocity $\langle \dot{x} \rangle$ as a function of the tilt $\mu$. We see that for 
weak noise, $\Omega=10^{-3}$, the current is vanishing inside the interval $(-1,1)$ and increases positively, respectively negatively,
for $\mu>1$, respectively $<-1$. The particle `rolls' in the direction of the tilt. For increasing $\Omega$ in the Gaussian noise,
the current assumes non-zero values inside the region where deterministically it should vanish: the noise generates tails that allow the particle
to `spill' outside the potential wells giving a non-vanishing probability that the particle rolls with the tilt
when deterministically it should be stable; hence the particle is metastable.

We also examine these same properties through the probability density solutions of the Gaussian Fokker-Planck equation
as insets in Fig.\ref{fig:gaussPDF}.
The left-most inset provides a case with stable fixed points ($\gamma=1$, $\mu=0.99$) and the right-most inset details a case with no fixed points ($\gamma=1$, $\mu=1.01$). Both insets are presented as 
so-called `wrapped probability density functions' (see for example \cite{Fish96}), namely a parametric plot where
\begin{eqnarray}
\{ x,y \} &=& \left\{ \left(1+{\cal P}_{st}^{(r)}(\phi) \right)\cos \phi,  \left(1+{\cal P}_{st}^{(r)}(\phi) \right)\sin \phi \right\}, \nonumber\\
&&\textrm{for  } \;\;\phi \in \mathbb{S}^1.
\label{parametric}
\end{eqnarray}
Note that in such plots the convention is that the positive horizontal axis represents $\phi=0$ (which would correspond to the vertical axis for a
density on the real line), peaks located clockwise from this are
in the positive direction and those anti-clockwise are in the negative.

For the leftmost inset in Fig.\ref{fig:gaussPDF}, for small $\Omega$ we see that the density is approximately concentrated at the deterministic position of the stable fixed point ($\phi = \rho + \sin^{-1}(\mu/\gamma) \approx - 0.07$, and thus the peak is oriented slightly below the horizontal axis). Moreover, as $\Omega$ increases we see that the probability density begins to smear around the circle, losing any discernable features after $\Omega = 1$. 
This displays the phenomenon of metastability in the wrapped densities: the particle is rolling with some probability so the probability density is distributed around the entire circle. 

The narrative is similar for the right-most inset, except the relative heights of the densities is significantly less - indicating that even for the weakest noise there is non-vanishing probability that the
particle is at other points around the circle because deterministically the particle will roll for $\mu>1$. Further increases in $\Omega$
distends the peak until it is uniformly distributed around the circle.

The wrapped densities also provide insight into the range of fluctuations. So, for $\mu=1.01$ the weakest noise $\Omega=10^{-3}$
has a density that is evenly distributed on either side of the horizontal axis, indicating that fluctuations about the average of zero are symmetrically
distributed. As $\Omega$ increases, the bulge moves clock-wise around from the horizontal axis indicating that fluctuations
are biased in the positive direction - the direction of the tilt. Similar properties are seen for $\mu=0.99$ but at stronger values of $\Omega$. 

\section{Analytic solution of the TFFP equation}
\subsection{Reduced density}
As with the Gaussian case, we expect that the steady state equivalent of Eq.(\ref{defining-FP}) with vanishing boundary conditions at the natural boundaries $x \rightarrow \pm \infty$ will be non-normalisable due to metastability. In order to ameliorate this situation we again consider a reduced density defined on $\mathbb{S}^1$ (Eq.(\ref{reduced})) for the steady state TFFP equation
\begin{equation}
 \left\{ \Omega \partial^{\alpha,\theta,\lambda}_x + \frac{\partial}{\partial x} V'(x)  \right\} {\cal P}^{(r)}_{st}(x) = 0
\label{stat-Levy}
\end{equation}
with boundary and normalisation conditions given by Eq.(\ref{boundandnorm}). 

Taking the Fourier transform of Eq.(\ref{stat-Levy}) we obtain
\begin{eqnarray}
\begin{split}
&\left\{ \Omega \Lambda(k) + ik \mu \right\}   {\widehat{\cal P}}^{(r)}_{st}(k) \\
&= \frac{k \gamma}{2} \left\{ e^{-i \rho} {\widehat{\cal P}}^{(r)}_{st}(k+1) - e^{i \rho} {\widehat{\cal P}}^{(r)}_{st}(k-1)    \right\}\\
\Rightarrow & \;\;\widehat{Q}(k+1) = - f_k  \widehat{Q}(k) +  \widehat{Q}(k-1),
\end{split}
\label{DefEQ}
\end{eqnarray}
where $\widehat{Q}(k) = e^{-i k \rho} {\widehat{\cal P}}^{(r)}_{st}(k)$ and
\begin{equation}
f_k = - \frac{2}{k \gamma}  \left\{ \Omega \Lambda(k) + ik \mu \right\} .
\end{equation}
Eq.(\ref{DefEQ}) represents a {\it linear} three-term recurrence relation defining the coefficients in the Fourier expansion.

Because of the periodic boundary conditions of ${\cal P}^{(r)}_{st}(x)$ on the finite interval, the Fourier variable $k$ only takes discrete values.
Specifically,
\begin{eqnarray*}
{\cal F} \left[{\cal P}^{(r)}_{st}(x)\right] = \sum^{\infty}_{n=-\infty}{\cal F} \left[ {\cal P}_{st}(x+2 \pi n)\right] \\
=  \widehat{{\cal P}}_{st}(k) \sum^{\infty}_{n=-\infty} e^{-i 2 \pi n k}\\
=  \widehat{{\cal P}}_{st}(k) \sum^{\infty}_{m=-\infty} \delta(k-m).
\end{eqnarray*}
Thus we need to solve the three term recurrence relation for coefficients $\widehat{Q}(k)$ in Eq.(\ref{DefEQ}) for $k \in \{\dots, -1,0,1,\dots\}$. Doing so, we can then construct the probability density via the discrete inverse Fourier transform
\begin{equation}
{\cal P}^{(r)}_{st}(x) = \frac{1}{2 \pi} \left\{ \widehat{Q}(0) + 2 \Re \sum^{\infty}_{n=1} e^{-i n (x-\rho)} \widehat{Q}(n) \right\}
\label{PDF}
\end{equation}
where $\widehat{Q}(0) = 1$ from the normalisation condition in Eq.(\ref{boundandnorm}).

Following Chap.9 of \cite{Risken89}, applying the transformations
\begin{equation*}
S_{k+1} = \frac{\widehat{Q}(k+1)}{\widehat{Q}(k)}
\end{equation*} 
the linear three term recurrence relation in Eq.(\ref{DefEQ}) becomes the following {\it non-linear} two term recurrence relation
\begin{equation}
S_k = \frac{1}{f_k + S_{k+1}}
\end{equation}
which can be solved iteratively using continued fractions
\begin{eqnarray*}
S_k = \cfrac{1}{f_k+\cfrac{1}{f_{k+1}+ S_{k+2}}} = \cfrac{1}{f_k+\cfrac{1}{f_{k+1}+\cfrac{1}{f_{ k+2} + \cfrac{1}{\ddots} }}}\\
\end{eqnarray*}
Moreover, applying the notation
\begin{equation*}
\mathbb{K}^{m}_{j=1}(a_j:b_j) = \cfrac{a_1}{b_1+\cfrac{a_2}{b_{2}+\cfrac{a_3}{\dots + \cfrac{a_m}{b_m} }}}
\end{equation*}
$S_k$ can be conveniently expressed by
\begin{equation}
S_k = \mathbb{K}^{\infty}_{j=k}(1:f_k).
\label{cont-notation}
\end{equation}

Calculating the expressions for $\{S_k, S_{k-1}, \dots, S_2,S_1\}$, we may then reconstruct the corresponding $\widehat{Q}(k)$ using
\begin{equation}
\widehat{Q}(k) = S_k S_{k-1} \dots S_2 S_1  \widehat{Q}(0)
\label{SolN}
\end{equation}
for insertion into Eq.(\ref{PDF}). Thus Eq.(\ref{PDF}) with coefficients given by Eq.(\ref{SolN}) is the solution to the TFFP equation, equivalent to 
Eq.(\ref{PDFGauss}) for Gaussian noise.

For numerical calculations of the reduced density we truncate the number of terms in Eq.(\ref{PDF}) to $n=\{1,\dots,1000\}$. Moreover, for the continued fractions $S_k$ in Eq.(\ref{cont-notation}) that form the density in Eq.(\ref{PDF}), we approximate these as $S_k = \mathbb{K}^{p}_{j=k}(1:f_k)$, where $k=\{1,\dots, 1000\}$, and $p =2000$.

\subsection{Average velocity}
Considering the expression for the the average velocity $\langle \dot{x} \rangle \equiv \frac{d}{dt} \langle x \rangle$ given in Eq.(\ref{aveGauss})
we may perform the following sequence of manipulations to re-express the expected current in terms of the characteristic function
for the tempered stable L{\'e}vy process:
\begin{eqnarray*}
\frac{d}{dt}\langle x \rangle & =& \frac{d}{dt}\langle x e^{i k x} \rangle|_{k \rightarrow 0}\\
&=& \frac{\partial}{\partial t} \left\{ -i \frac{\partial}{\partial k}\widehat{ {\cal P}}(k,t)  \right\}|_{k \rightarrow 0}\\
&=& - i \frac{\partial}{\partial k} \left\{ \frac{\partial}{\partial t} \widehat{ {\cal P}}(k,t) \right\}  |_{k \rightarrow 0}\\
&=& - i \frac{\partial}{\partial k} \left\{ \Omega \Lambda(k)\widehat{ {\cal P}}(k,t) - i k {\cal F} \left[  V'(x) {\cal P}(x,t) \right] \right\}  |_{k \rightarrow 0}\\
&=& -i \Omega \frac{d}{dk} \Lambda(k)|_{k \rightarrow 0} - {\cal F} \left[  V'(x) {\cal P}(x,t) \right]  |_{k \rightarrow 0}.
\end{eqnarray*}
From Eq.(\ref{lambda})
\begin{equation*}
-i \Omega \frac{d}{dk} \Lambda(k)|_{k \rightarrow 0} = \left\{  \begin{array}{cl}
-\frac{\Omega \alpha \theta \lambda^{\alpha -1}}{\cos \left( \frac{\pi \alpha}{2} \right)} & 0 <\alpha < 1 \\
0  & 1 < \alpha < 2
\end{array} \right.
\end{equation*}
and from Eqs.(\ref{ratchetpot}) and (\ref{PDF})
\begin{equation*}
- {\cal F} \left[  V'(x) {\cal P}(x,t) \right]  |_{k \rightarrow 0} = \mu - 2 \pi \gamma \Im \left\{ \widehat{Q}(1) \right\}.
\end{equation*}
Hence the expected value of the velocity may be expressed as
\begin{equation}
\langle \dot{x} \rangle  =  \left\{  \begin{array}{cl}
-\frac{\Omega \alpha \theta \lambda^{\alpha -1}}{\cos \left( \frac{\pi \alpha}{2} \right)} + \mu -  2 \pi \gamma \Im \left\{ \widehat{Q}(1) \right\} & 0 <\alpha < 1 \\
\mu -  2 \pi \gamma \Im \left\{ \widehat{Q}(1) \right\}  & 1 < \alpha < 2.
\end{array} \right.
\label{Levyvel}
\end{equation}
Eq.(\ref{Levyvel}) with $ \widehat{Q}(1)$ given by Eq.(\ref{SolN}) is the tempered-stable equivalent of Eq.(\ref{aveGauss}) for Gaussian noise.

For the corresponding numerical calculations of the average velocity in Eq.(\ref{Levyvel}), only the first continued fraction coefficient $\widehat{Q}(1) = S_1 \widehat{Q}(0)$ is involved. For numerical purposes we again truncate $S_1$  as $ S_1 = \mathbb{K}^{p}_{j=1}(1:f_k)$. For $\alpha>1$ we apply $p=2\times10^4$. However, for $\alpha < 1$ we have found it necessary to set $p=4\times 10^4$ to obtain sufficiently smooth plots of the average velocity.

\section{Examples: Current stabilisation and reversal}
As for the Gaussian case, we fix the amplitude $\gamma=1$ so that the deterministic threshold for instability is
$\mu=\pm 1$. We also fix the diffusivity constant at $\Omega=0.1$, where the Gaussian case in Fig.\ref{fig:gaussPDF}
shows diffusion even for $\mu<1$ due to metastability, and $\rho=-1.5$ as before.
We first examine the average velocity and wrapped densities for selective values of $\alpha$ and $\lambda$.

\begin{figure*}
\includegraphics[height=7cm]{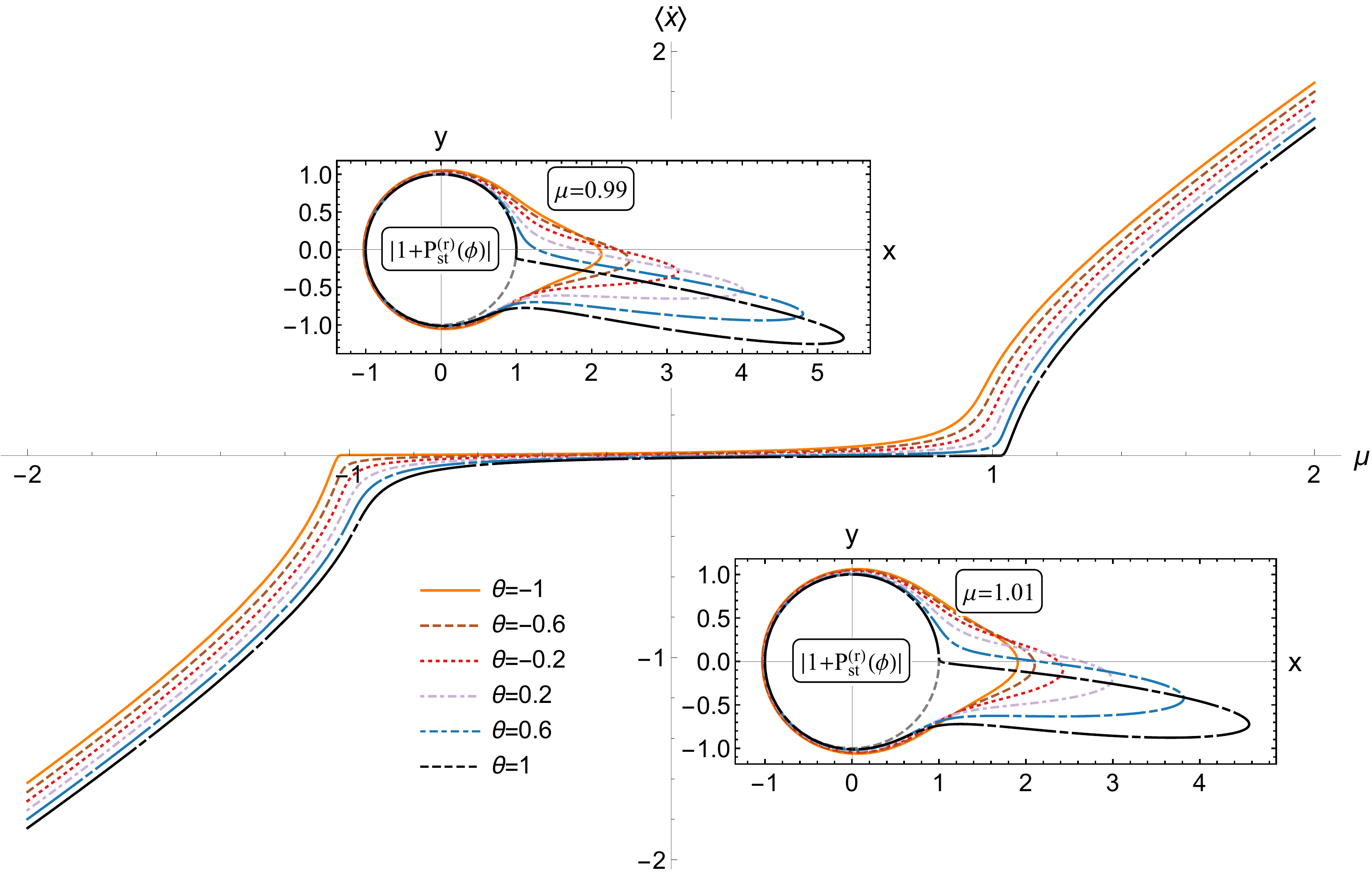}
\caption{Examples of Eq.(\ref{PDF}) (insets) and Eq.(\ref{Levyvel}) (main) for $\rho=-1.5$, $\gamma=1$, $\alpha=0.5$, $\lambda=0.5$, $\Omega=0.1$ and various values of the asymmetry $\theta$. The left-most inset details a case with stable fixed points in the potential ($\mu=0.99$), the right-most inset details a case with no fixed points in the potential ($\mu=1.01$). Both insets are given as a parametric plot governed by Eq.(\ref{parametric}). }
\label{fig:alpha05}
\end{figure*}

With $\alpha=0.5$ we expect for the stable noise case ($\lambda=0)$ very heavy tails which will lead to quite diffuse densities.
To identify structure we therefore choose for this $\alpha$ relatively large tempering, namely $\lambda=0.5$.
The analogue to  Fig.\ref{fig:gaussPDF} is shown in  Fig.\ref{fig:alpha05}, where again the average velocity is shown as a function of $\mu$
for different values now of skew $\theta$, and insets show the wrapped densities for two the choices of $\mu$ above and below
the threshold $\mu=1$. 

The signature feature of Fig.\ref{fig:alpha05} is the behaviour around $\mu=1$: we see that for $\theta=1$ (black curve)
the average velocity is zero for a small range of values $1< \mu\leq 1.05$; near $\mu=-1$ the same behaviour recurs for $\theta=-1$.  Once the skew decreases, $\theta<1$ for $1< \mu\leq 1.05$, and increases $\theta>-1$ for$-1.05\leq \mu<-1 $, the average velocity becomes non-zero
in these narrow regions. In other words, one-sided tempered stable noise
anti-aligned to the tilt may stabilise the particle on average in circumstances where deterministically it should be unstable (namely there is no well).
This is {\it stochastic current stabilisation}.

These properties are reflected in the wrapped densities in the inset. The left-most inset of Fig.\ref{fig:alpha05}, for $\mu=0.99$
shows strongly peaked densities just as for the Gaussian case. However, the right-most inset with $\mu=1.01$
also shows a strongly peaked density for $\theta=1$ where the analogous Gaussian case of Fig.\ref{fig:gaussPDF} shows
diffusion around the circle (particularly for the corresponding value of $\Omega$). Observe how for both cases of $\mu$, 
the curve for the one-sided density with $\theta=1$ joins the circle sharply on one side (anti-clockwise from the peak) and more smoothly
on the other (clockwise from the peak). This manifests the skew for this case.

\begin{figure*}
\includegraphics[height=7cm]{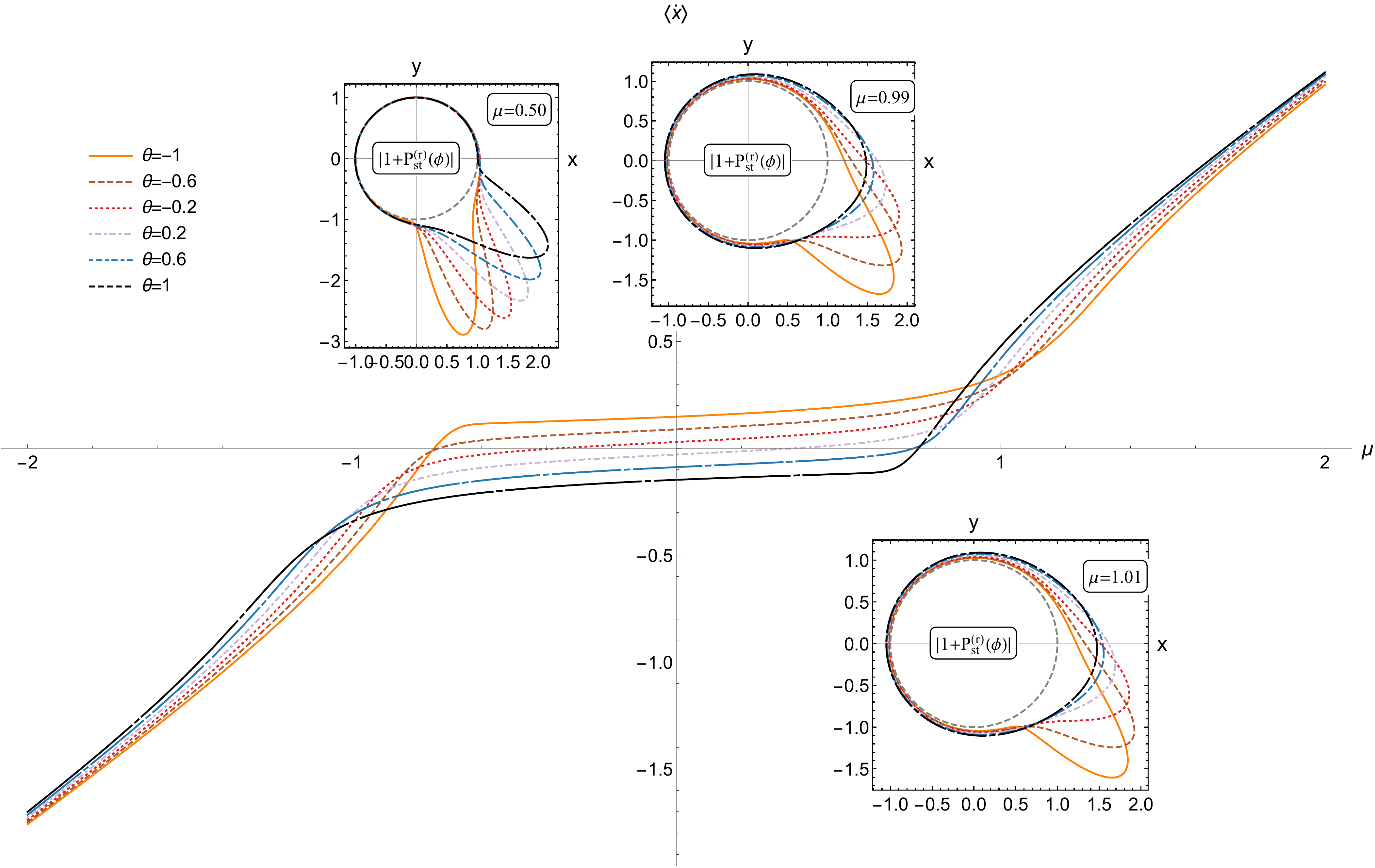}
\caption{Examples of Eq.(\ref{PDF}) (insets) and Eq.(\ref{Levyvel}) (main) for $\rho=-1.5$, $\gamma=1$, $\alpha=1.25$, $\lambda=0.001$, $\Omega=0.1$ and various values of the asymmetry $\theta$. The left-most inset details a case with stable fixed points in the potential ($\mu=0.5$), the
middle inset is $\mu=0.99$ and the right-most inset details a case with no fixed points in the potential ($\mu=1.01$). The insets are given as a parametric plot governed by Eq.(\ref{parametric}). }
\label{fig:alpha125}
\end{figure*}

We now chose a contrasting case with $\alpha=1.25$, greater than one but still significantly far from Gaussianity, and $\lambda=0.001$ 
which is close to the stable limit. The corresponding average velocity and wrapped densities are shown in Fig.\ref{fig:alpha125},
but now for three choices of $\mu$.
Now for $-1<\mu<1$ the average velocity no longer vanishes in general except for specific values of $\theta$ at specific values of $\mu$;
at such values the skew exactly balances against the tilt. 
However, unusually, there are regions where the sign of $\langle \dot{x} \rangle$ is {\it opposite} to that of $\mu$ for certain ranges of skew $\theta$.
For example, for $\theta=1$ (black curve) and $0<\mu<0.7$,  $\langle \dot{x} \rangle<0$ the particle is propagating in the negative direction
(to the left) even though skew and tilt are positive. For $\theta=-1$ (orange curve) and $-0.7<\mu<0$,  $\langle \dot{x} \rangle>0$ so that the particle is propagating in the positive direction
(to the right) even though skew and tilt are negative. This is the phenomenon of current reversal. 

Examining the wrapped densities, shown in the insets, we see strongly oriented and sharp densities for $\mu=0.5$
across the range of $\theta$,
while for $\mu=0.99, 1.01$ the densities are mostly diffuse for $\theta>0$, most strongly for $\theta=1$. These assist, to a degree, in understanding the counter-intuitive
current reversal. Specifically, we observe that for the one-sided case $\theta=-1$ (orange curve) the heavy tail is in the anti-clockwise direction
while the peak is oriented clockwise from the positive horizontal direction. Recalling that all of the cases of $\theta$ correspond to the same mean,
we observe that the mode of the distribution shifts further and further clockwise from the axis. This indicates that the heavy tail is in
the anti-clockwise direction (in contrast to the $\theta=1$ case). Significantly, for $\alpha>1$ the {\it mode of the distribution
lies in the opposite direction from the heavy tail} \cite{KallRob2017}.
The wrapped densities, particularly with the distortions of plotting on a circle, do not convey well the significance of the heavy tail;
in \cite{KallRob2017} the role of the tail was best represented through Quartile-Quartile plots.
Nevertheless, in the case of such small tempering given a heavy tail in the negative direction for $\theta<0$, we obtain a probability mass
in the positive direction. So there is a dominance of small jumps in the positive direction. Thus, with
small tempering and even subject to positive tilt, there is a net drift in the positive direction. The current reversal arises from this.
(We only caution that the threshold for this behaviour cannot immediately be read off the wrapped density plots since they
represent the density of $x$ rather than the current density.)
A similar phenomenon was observed in the Kuramoto model in \cite{KallRob2017}, a point to which we shall return in the final section.

\begin{figure*}
\includegraphics[width=18cm]{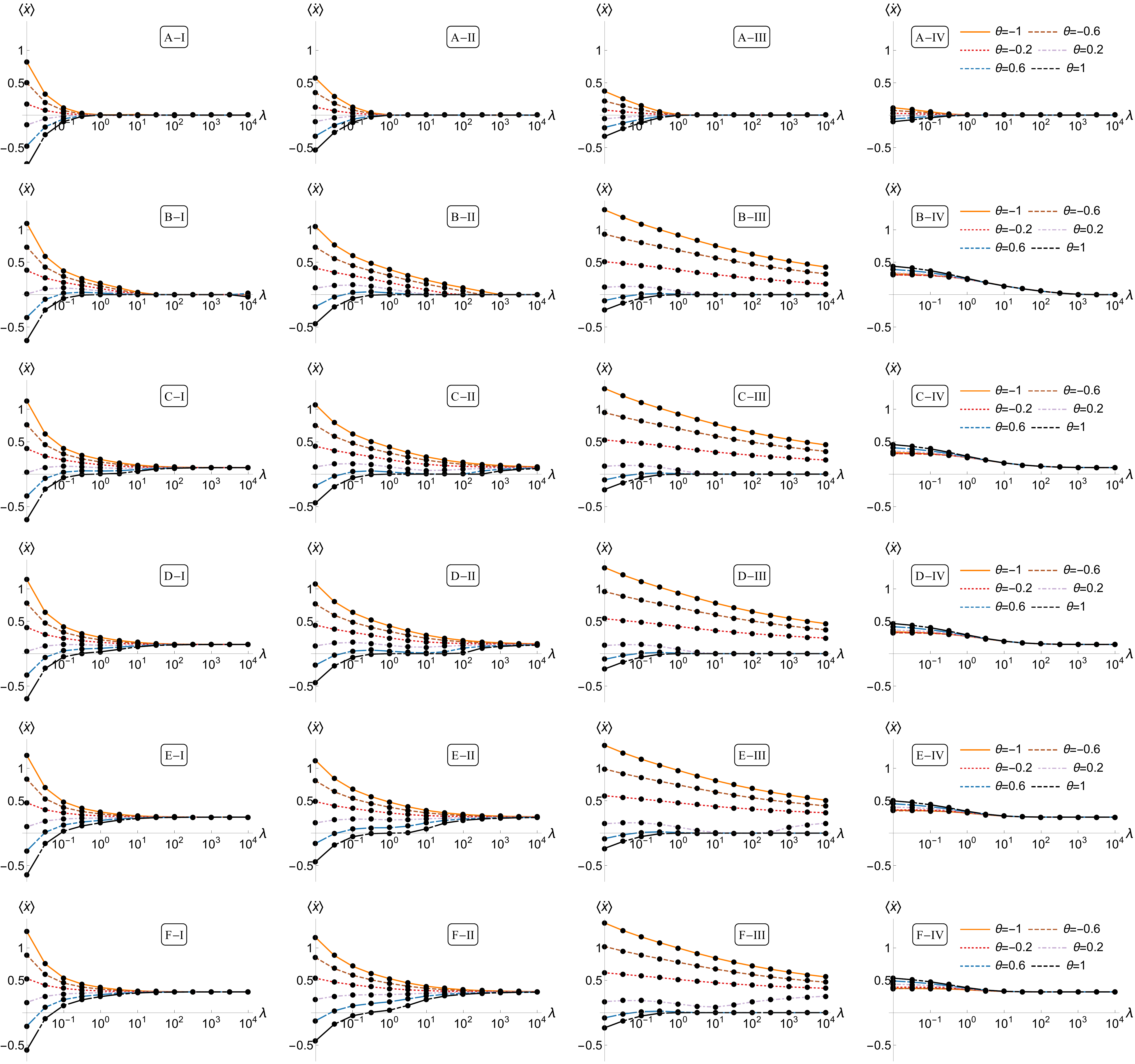}
\caption{Examples of Eq.(\ref{Levyvel}) for $\rho=-1.5$, $\gamma=1$, $\Omega=0.1$. The tempering $\lambda$ is being varied on the horizontal axis, with various values of the asymmetry $\theta$. Down the rows (A-F) the tilt $\mu$ is varied, with the values $\mu = \{ 0.1, 0.995, 1.005, 1.01, 1.03, 1.05\}$ being applied from top to bottom row. Across the columns (I-IV) the fractional power $\alpha$ is varied, with the values 
$\alpha = \{ 0.25,0.55,0.85,1.25 \}$ applied from left-most to right-most column.}
\label{fig:currentstudy}
\end{figure*}

Having examined two selected cases, we now conduct a more general scan across a range of $\lambda$ for representative values
of $\alpha$ below and above one and different skew values $\theta$. This is shown in Fig.\ref{fig:currentstudy}.
These plots should be compared with Fig.15 in \cite{KullCast2012} for
tempered stable noise in a (different to ours) ratchet potential without tilt $(\mu=0$). We choose $\mu$ quite far from this regime, and scan in the
vicinity of $\mu=1$, namely $0.995\leq \mu \leq 1.05$ and $\alpha=0.25,0.55,0.85, 1.25$. But for some comparison we also
provide the result for $\mu=0.1$. As $\mu>0$ for
all these cases, current reversal corresponds to negative values of $\langle \dot{x} \rangle$ in these plots.

In fact, we observe that current reversal is a generic feature both below the deterministic threshold of $\mu=1$ and above: starting
from the top left and scanning across to higher values of $\alpha$, and scanning down with increasing
increments in $\mu$. The case shown in \cite{KullCast2012}, where they choose $\alpha=1.5$
and two values of $\theta$, shows a current starting negative for small $\lambda$, crossing zero to positive values
and then converging to zero; this is hidden in a plot such as panel A-IV in Fig.\ref{fig:currentstudy}. 
For larger tilt $\mu$ such behaviour moves to $\alpha<1$. For example, panel B-I shows curves for $\theta \approx 0.6$
crossing from negative to positive and then converging to zero.
This current reversal occurs not only for the most extreme skew $\theta=1$  but closer to symmetric noise.

We see that for $\mu<1$ and large $\lambda$ the average velocity tends to zero (top two rows); for $\mu>1$ the asymptotic limit
is non-zero and positive (third row and below), which corresponds to the deterministic value as tempering suppresses all noise.
We also observe that the current reversal ceases above $\alpha=1$ in this region of $\mu$ - recall that in Fig.\ref{fig:alpha125}
the reversal occurs for $\mu<0.7$.

Within the regimes of current reversal there are always discrete values where $\langle \dot{x} \rangle=0$.
However, we also see for $\mu>1$ and $\alpha<1$ regimes where the average velocity {\it vanishes across a continuum}
of $\lambda$ values before increasing and converging to the deterministic limit - this is most distinct in panel C-II.
Thus current stablisation may be sustained across a broad range of $\lambda$ before tempering dampens the noise completely.
These demonstrate that current stablisation is linked naturally to current reversal - but the stablisation over a range of $\lambda$ may
not be intuitively expected.

Moving to the lower rows we observe that the range of $\lambda$ over which stabilisation occurs shrinks until it becomes only 
a discrete case of $\lambda$ except for pure one-sided noise with $\theta=1$. For the cases in panels E-III and F-III 
with $\theta=1$ the average velocity will converge to its non-zero value at $\lambda$ values beyond those plotted here.
\\

\section{Conclusions and Discussion}
We have solved the Fokker-Planck equation for a particle in a one-dimensional tilted ratchet potential under
tempered stable L{\'e}vy noise and have observed both phenomena of current stabilisation and reversal,
particularly in regimes where deterministic or Gaussian considerations would show quite different behaviour.

The essence of the mechanism is the interplay between probability mass around mode of the underlying noise distributions and the heavy tails and how it shifts as
$\alpha,\lambda$ are varied. Specifically, for $\alpha>1$ the mode is typically in the opposite direction from the tail for asymmetric noise
as discussed in \cite{KallRob2017}.
This leads to a drift in a direction corresponding to the sign of the mode, which manifests as current reversal for the particle in the tilted ratchet.
When $\alpha<1$, due to the induced drift through the form of the noise characteristic function, the mode and heavy tails are in the same direction
but with increased mass around the mode and heavier tails. Thus, with tempering, both long range and intermediate range jumps are suppressed 
so that the effective drift may even drop down to zero with increased skew in the noise. Moreover, tempering moderates current reversal
until it assumes the form of current stabilisation. Thus the particular cases of current reversal in \cite{KullCast2012}
are a special instance of a phenomenon in the presence of tilt in the potential.

As alluded in the introduction, for us the interest in this phenomena arises from our work on the Kuramoto model or related forms.
In \cite{KallRob2017} and \cite{KallRob2017b} we have observed a phenomenon of oscillators of zero native frequency
nevertheless synchronising to a non-zero net frequency, a drift, that shifts in sign according to the interplay of $\alpha$ and $\lambda$.
While general observations of the position of the mode and the heaviness of the tail in those cases provide an heuristic explanation of
this phenomena, we argue that the model solved here may provide a deeper insight into this behaviour.
Future applications of this idea lie in its exploitation for stochastic control of synchronisation phenomena.

\end{document}